\documentstyle[prc,aps,twocolumn,psfig]{revtex}
\def\f{\frac}
\begin{document} 
\twocolumn[\hsize\textwidth\columnwidth\hsize\csname @twocolumnfalse\endcsname
\flushright{DOE/ER/40561-298-INT96-21-003}
\flushright{CU-TP-794}
\title{Frame Dependence of Parton Cascade results}
\author{Bin Zhang}
\address{Department of Physics, Columbia University, New York, NY 10027\\
and\\
INT, University of  Washington, Seatle, WA 98195-1550\\
}
\author{Yang Pang}
\address{Department of Physics, Columbia University, New York, NY 10027}
%\date{\today}
\date{6th November 1996}
\maketitle

\begin{abstract}
Frame dependence of parton cascade results is studied for different schemes of
doing cascade simulations. We show that different schemes do not always agree
and results may have strong frame dependence. When the interaction range is on
the order of mean free path and the collisions are done in the two parton
center of mass frame, the results are not sensitive to the global frame that
the collisions are ordered. Effects of different differential cross sections
are also discussed.

\vspace{0.5in}
\end{abstract}
]

\section{INTRODUCTION}
Cascade models have been widely used to simulate Relativistic and
Ultra-Relativistic Nucleus-Nucleus collisions \cite{ypang1}. They can be used
to simulate systems that are not in thermal equilibrium and they incooperate
the effect of finite mean free path automatically. 

Causality violation is inherent in cascade simulations due to the geometrical
interpretation of 
the cross section, i.e., particles scatter when their closest distance is
smaller than $\sqrt{\f{\sigma}{\pi}}$ ($\sigma$ is the  scattering cross
section). One problem is that the information travels across
$\sqrt{\f{\sigma}{\pi}}$ at a time which is either the fixed time step or the
mean free path. This may lead to a speed of information faster than the speed
of light and may cause 
superluminous shock waves. 
Another problem is that different choices of doing 
collisions, i.e., different
collision schemes, may lead to different collision orderings and hence
different physical predictions. 

Many groups have studied the causality problem for energies below RHIC energy
\cite{peter1} and attempts have been made to reduce the problem. 
For Ultra-relativistic
heavy ion collisions at RHIC energies and beyond, causality violation may be
very serious due to the relatively small mean free path which may be on the
order or 
even much less than the interaction range.  The superluminous signals have been
studied \cite{kortmeyer1} in cascade simulations of pA collisions. In this
paper, we study the effect of collision ordering on macroscopic variables by
comparing the results of different collision schemes. A parton cascade
code, GPC (Generic Parton Cascade), we developed recently \cite{zhang1} has
been used in this study.

First, we describe the initial conditions and different collision schemes. Then
we compare the results from different schemes with a discussion of the effects
of different differential cross sections followed by the conclusions.

\section{INITIAL CONDITIONS AND COLLISION SCHEMES}

To study frame dependence of cascade results at RHIC energies, we prepare a 
system of gluons similar to the minijet gluon system that is going
to be produced at RHIC. 4000 gluons are uniformly distributed in
the $-5$ to $5$ space time rapidity range. Initially, they get local thermal
equilibrium at temperature $500\;MeV$. They occupy a transverse disk of radius
$5$ fm at $t=0$ and are formed at longitudinal proper time $\tau_0=0.1\;fm$
(i.e., the 
formation time for a particular particle $i$ is $t_i=\tau_0\times
cosh(\eta_i)$, $z_i=\tau_0\times sinh(\eta_i)$).  

We will compare results in different frames of ordering the collisions for the
following schemes of doing cascade: 

(a) Two gluons collide when their closest approach distance in the two particle
    center of mass frame is smaller than $\sqrt{\f{\sigma}{\pi}}$ ($\sigma$ is
    the  scattering cross section). The collision space point is chosen to be
    the  midpoint of the two particles in the two-body center of mass frame at
    their  closest distance, the collisions are ordered in a global frame,
    either the  collider lab frame or the target frame (they differ by $6$
    unit of rapidity  boost);

(b) differs from (a) in that the collision space point for a particle (not for
    a collision) is the position of the particle in the two-body center of mass
    frame. So, in the global frame, for a particular parton parton collision,
    each parton will have its own time of scattering. The collisions are
    ordered according to the average time of the two scattering times in the
    global frame;

(c) differs from (b) in that the collisions are ordered according to the
    earlier time of the two scattering times;

(d) two partons collide when their closest distance is smaller than
    $\sqrt{\f{\sigma}{\pi}}$ in the global frame. The collisions are ordered
    according to the collision time in the global frame.

Different schemes specify different parton collision frame and different
collision ordering time.

The system propagates from one collision to the next in the global frame
instead of from one time step to the next.

Partons are all on mass shell. Rutherford cross section regulated by a
screening mass, $\f{d\sigma}{d\hat{t}}=\f{9 \pi
\alpha_S^2}{2(\hat{t}-\mu^2)^2}$, 
is taken to generate the scattering angular distribution. In the above formula,
$t=(p_1-p_2)^2$ in which $p_1$ and $p_2$ are 4 momenta of one particle before
and after the collision. The total cross
section for parton collision is taken as: 
\begin{equation}
\sigma=\f{9 \pi \alpha_S^2}{2 \mu^2}.
\end{equation}
We set $\alpha_S=\sqrt{\f{2}{9}}\approx 0.47$

Numerically \cite{zhang1}, we divide the space into cells. We store only the
next collision time, partner and next particle in the same cell in an
interaction list to 
save memory and memory manipulation. The interaction list is only updated
locally, i.e., only the cell where the collision happens and its neighboring
cells are updated. This reduces the number of checks and significantly
increases the speed of the parton cascade code. We can get one $6,000$
collisions event in around 10 minutes. Our results have been checked by taking
away the cells to make sure they are consistent with the ordinary calculations
without space divisions. At the present stage, there are only 2
to 2 scatterings. We are working on including the radiation processes into the GPC.

\section{Results for different schemes}

The $\f{dN}{dy}$, $\f{dE_T}{dy}$ and $\f{dN}{dp_T}$ distribution of partons are
closely related to the hadronic observables. In this section, we
are going to study parton cascade induced changes of these
distributions. First, we look at the case when the mean free path is on the
order of the interaction range $\sqrt{\f{\sigma}{\pi}}$; then the case when the
mean free 
path is much smaller. Further, we study the dependence of the results on the
differential cross section, i.e., the angular distribution of outgoing
particles from a collision. Finally, a discussion on the number of total
collisions vs. the number of non-causal collisions and effects of a small
particle mass.

The mean free path $l\sim\f{1}{n\;\sigma}$ in which $n$ is the parton number
density and the cross section $\sigma\sim \f{\pi}{\mu^2}$. 
The density $n$ can be estimated through:
\[n=\f{1}{\pi\;R^2 t}\f{dN}{dy}.\]
In our case, $R=5fm$, $\f{dN}{dy}=400$ and $t\sim0.2fm$ taken into account the
spread of rapidity around the space time rapidity. We get $n\sim 25/fm^3$. 

The interaction
range $\sqrt{\f{\sigma}{\pi}}\sim \f{1}{\mu}$.

To have $l>\sqrt{\f{\sigma}{\pi}}$, we need to have $\mu>4.3\;fm^{-1}\approx
0.85\;GeV$. This is a condition for isolated 2 body collisions that is
generally violated. Ideally we also need many particles within the Debye
screening scale:
\[N_D=n(\f{4}{3}\f{\pi}{\mu^3})\gg 1.\]
For $\mu=3\;fm^{-1}$, $N_D\approx\f{100}{3^3}\approx 4$.

We showed below even though the isolated 2 body collision condition is barely
satisfied, in the situation with strong inside-outside correlations like the
system evolving from the initial conditions we specified above, a sensible
scheme can still produce reasonable results.

Fig.~1, Fig.~2 and Fig.~3 give the $\f{dN}{dy}$, $\f{dE_T}{dy}$ and
$\f{dN}{dp_T}$ distributions with cross section $\sigma=\f{\pi}{9}\;fm^2$
($\alpha=0.47$ and $\mu=3\;fm^{-1}$) respectively. In plotting the target frame
ordering case, we make a shift so that the central rapidity always has
rapidity $0$.

After evolution,
$\f{dN}{dy}$ is almost the same as the initial. $\f{dE_T}{dy}$ has a drop from
initial value $\sim\;385\;GeV$ to around $335\; GeV$. This shows the
longitudinal work \cite {ruus1} done by the effective pressure. The absolute
value of slope 
of $\f{dN}{dp_T}$ log curve is increasing which clearly shows the effective
cooling of the system. The results do not depend on the frame we use to order
the collisions. 

Scheme (b) and (c) give similar results.

A close look at the average number of collisions per event shows that for
scheme (a) we get $\sim\;6900$ collisions, for (b) $6800$ and for (c) $6200$
when collision order is in the collider lab frame. In the target frame case,
each gets several hundred less.

\begin{figure}[h]
\vspace{1.5cm}
\hspace{0.01cm}
\psfig{figure=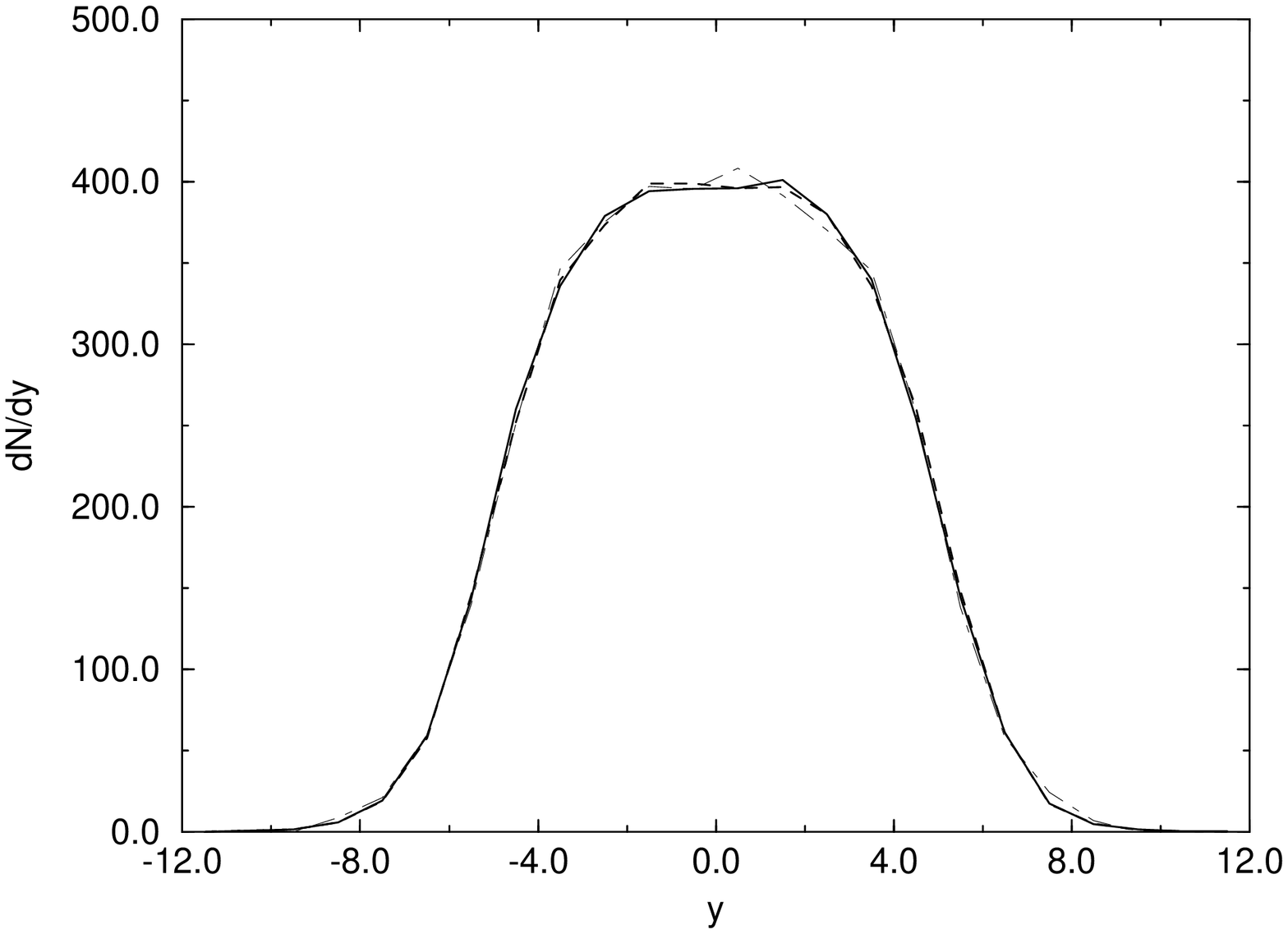,height=2.1in,width=1.9in,angle=-90} 
\vspace{-0.5cm}
\caption{
20 event averaged $\f{dN}{dy}$ distribution for scheme (a) with $\alpha=0.47$
and $\mu=3\;fm^{-1}$. The dot-dashed curve is the initial distribution for
reference; the solid curve is for the case that collisions are ordered in the
collider center of mass frame; for the dashed curve, the collisions are ordered
in the target frame.
}
\end{figure}

\begin{figure}[h]
\vspace{1.5cm}
\hspace{0.01cm}
\psfig{figure=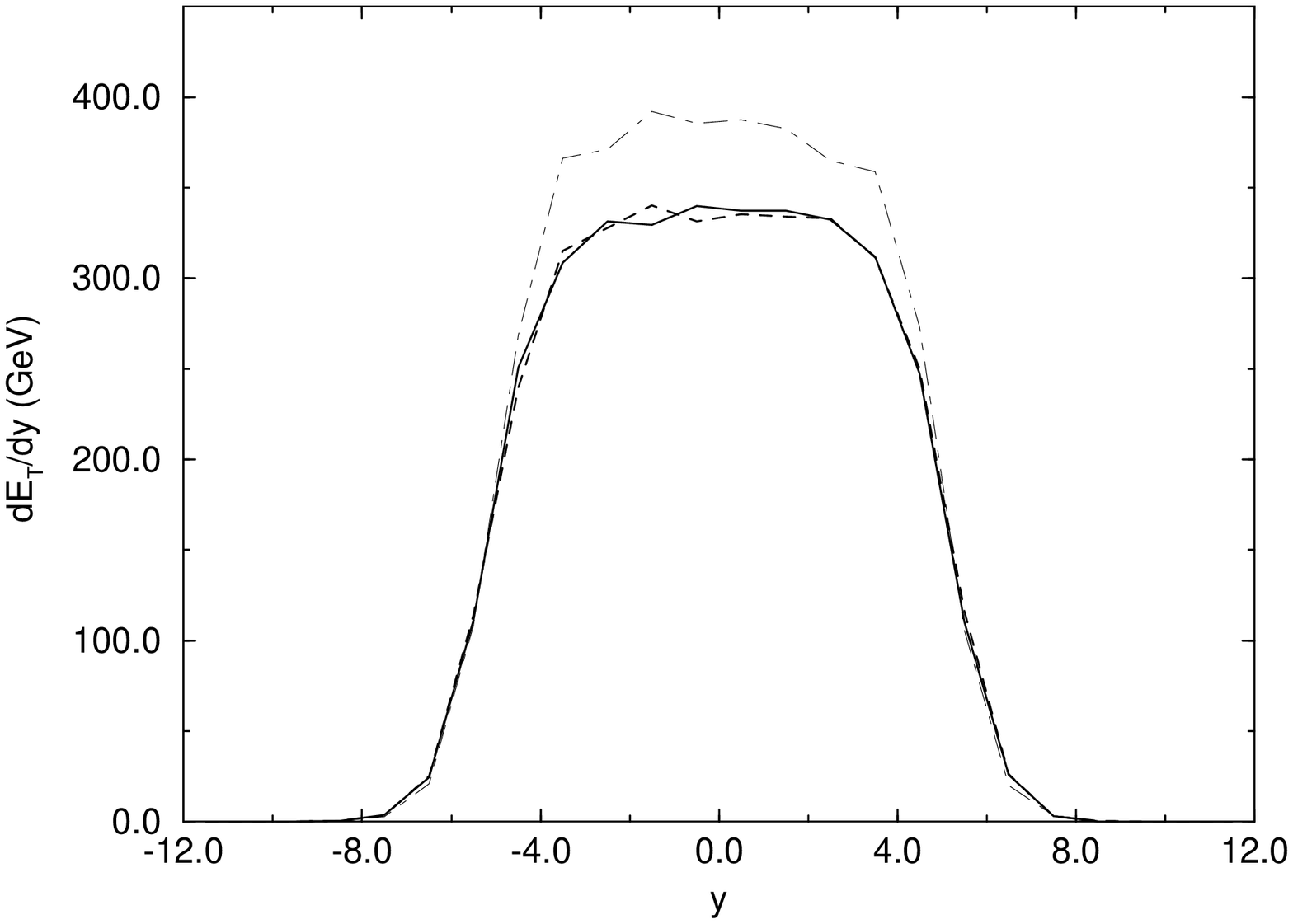,height=2.1in,width=1.9in,angle=-90} 
\vspace{-0.5cm}
\caption{
20 event averaged $\f{dE_T}{dy}$ distribution for scheme (a) with $\alpha=0.47$
and $\mu=3\;fm^{-1}$. The dot-dashed, solid and dashed curves are for the
initial, collider frame ordering and target frame ordering respectively.
}
\end{figure}

\begin{figure}[h]
\vspace{1.5cm}
\hspace{0.01cm}
\psfig{figure=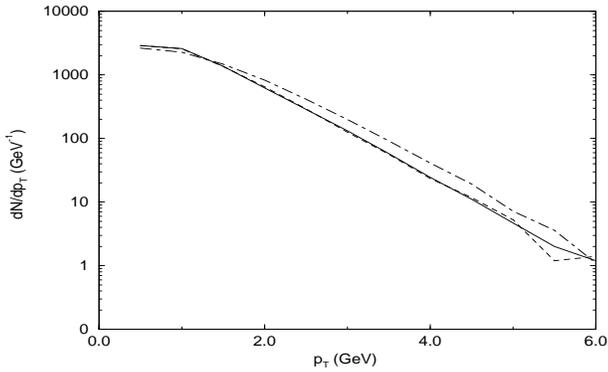,height=2.1in,width=1.9in,angle=-90} 
\vspace{-0.5cm}
\caption{
20 event averaged $\f{dN}{dp_T}$ distribution for scheme (a) with $\alpha=0.47$
and $\mu=3\;fm^{-1}$. The dot-dashed, solid and dashed curves are for the
initial, collider frame ordering and target frame ordering respectively.
}
\end{figure}

Fig.~4, Fig.~5 and Fig.~6 give the results for scheme (d). Strong frame
dependence is observed in this case. From Fig.~4 and 5, we see that a
boost of the system by 6 unit of rapidity (the original $0$ rapidity goes to
$-6$ before shifting back to $0$) makes a 
peak which is shifted from the central along the boost by 2 unit of
rapidity. We use a much smaller $\sigma=\f{\pi}{25}\;fm^2\approx 1mb$ cross
section because in scheme (d), we get much more
collisions even with a smaller cross section. In the collider frame ordering
case, there are around $22,000$ collisions per event; in the target frame
ordering case, we get $97,000$!

The frame dependence of scheme (d) is due to the rapidity correlation
introduced by taking the closest distance in the global frame. It neglects
the flow velocity and increases the density of the particles. This is also seen
in the huge increase of number of collisions with smaller cross section
comparing to the other schemes. Scheme (a), (b) and (c) take into account of
the flow by looking at the closest distance in the 2 particle center of mass
frame. This corresponds to the time in the 2 particle center of mass frame when
the 2 discs meet. 

\begin{figure}[h]
\vspace{1.5cm}
\hspace{0.01cm}
\psfig{figure=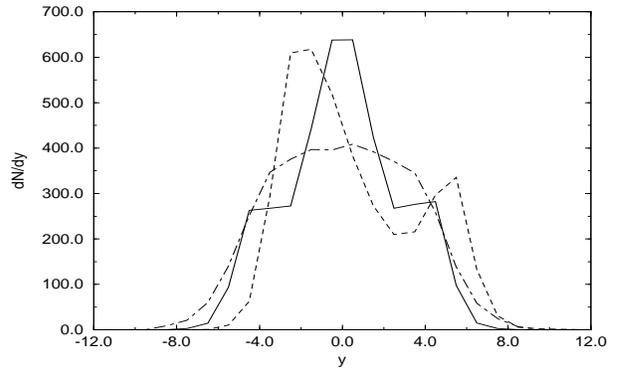,height=2.1in,width=1.9in,angle=-90} 
\vspace{-0.5cm}
\caption{
5 event averaged $\f{dN}{dy}$ distribution for scheme (d) with $\alpha=0.47$
and $\mu=5\;fm^{-1}$. The dot-dashed, solid and dashed curves are for the
initial, collider frame ordering and target frame ordering respectively.
}
\end{figure}

\begin{figure}[h]
\vspace{1.5cm}
\hspace{0.01cm}
\psfig{figure=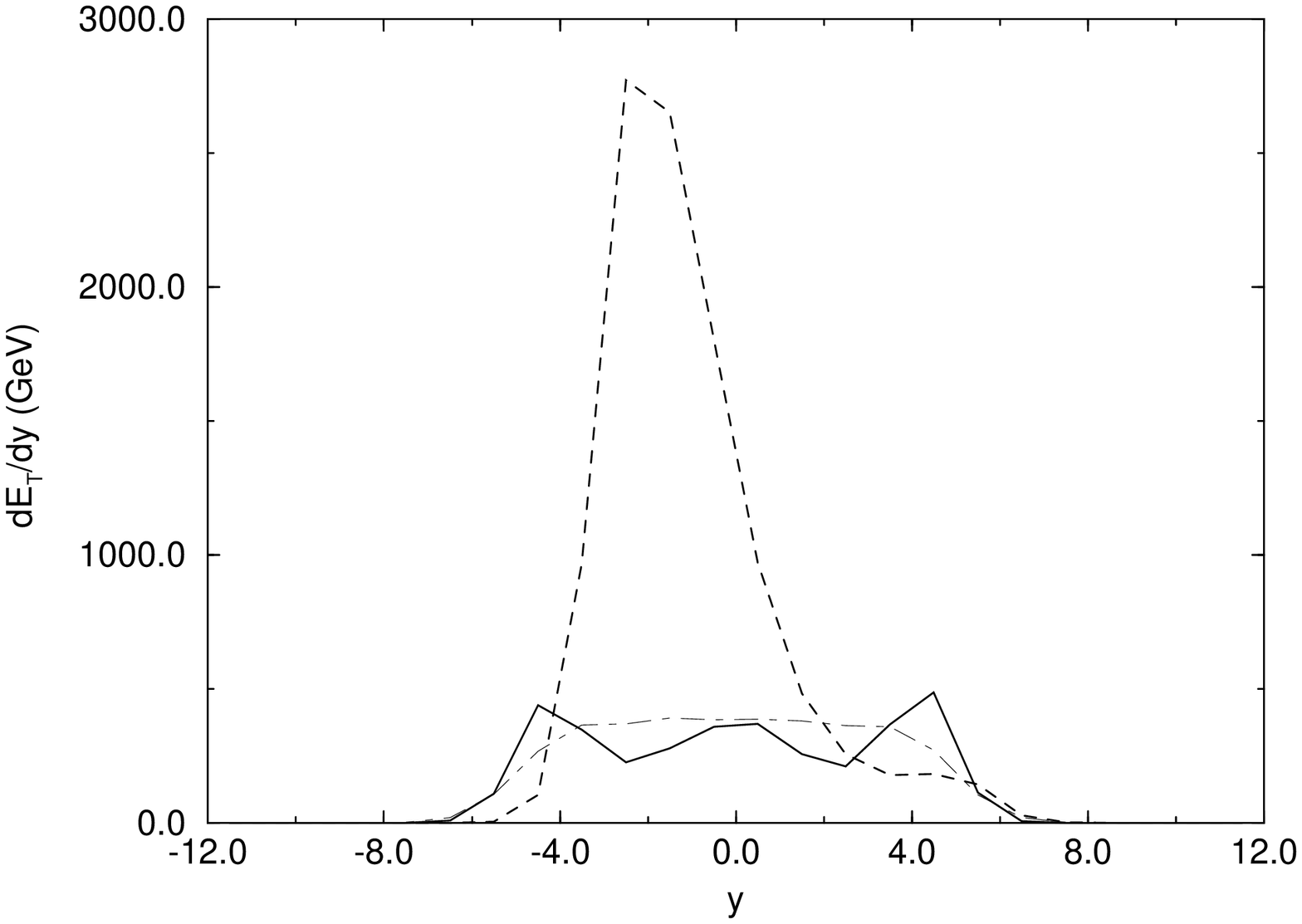,height=2.1in,width=1.9in,angle=-90} 
\vspace{-0.5cm}
\caption{
5 event averaged $\f{dE_T}{dy}$ distribution for scheme (d) with $\alpha=0.47$
and $\mu=5\;fm^{-1}$. The dot-dashed, solid and dashed curves are for the
initial, collider frame ordering and target frame ordering respectively.
}
\end{figure}

\begin{figure}[h]
\vspace{1.5cm}
\hspace{0.01cm}
\psfig{figure=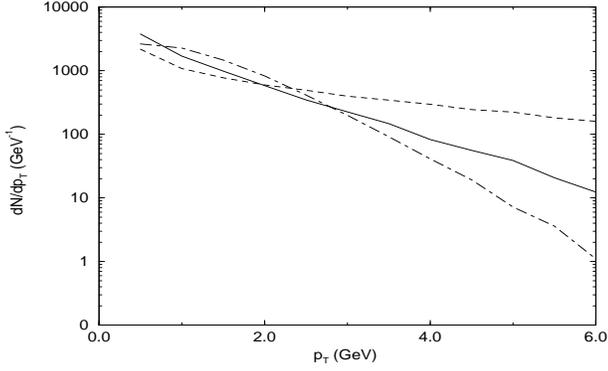,height=2.1in,width=1.9in,angle=-90} 
\vspace{-0.5cm}
\caption{
5 event averaged $\f{dN}{dp_T}$ distribution for scheme (d) with $\alpha=0.47$
and $\mu=5\;fm^{-1}$. The dot-dashed, solid and dashed curves are for the
initial, collider frame ordering and target frame ordering respectively.
}
\end{figure}

Fig.~7, Fig.~8 and Fig.~9 give results for $\sigma=\pi\;fm^2$. Now the
interaction range is $1fm$ and the mean free path is initially
$\f{1}{25\pi}\sim 0.01fm$. For
$\alpha=0.47$ and $\mu=1\;fm^{-1}$, the $dN/dy$ distribution has a dip in the
middle for the collider frame ordering case, while the target frame ordering
case doesn't have one. The $dE_T/dy$ central rapidity plateau height is almost
the same as the for $\alpha=0.47$ and $\mu=3\;fm^{-1}$ case. This is because of
the effective screening mass is used to regulate the forward scattering cross
section and larger cross section or more forward collisions does not lead to
more collective
work. This is clearly shown when we vary the differential cross section by
change the $\alpha$ from $0.47$ to $1.41$ and $\mu$ from $1\;fm^{-1}$ to
$3\;fm^{-1}$. As shown in Fig.~8, the collider frame ordering case has a clear
drop at 
rapidity from $-4$ to $-2$ and from $2$ to $4$. In the middle, there is a
peak. The target frame
ordering has a lower plateau. The $dE_T/dy$ is frame dependent now. The
$dN/dp_T$ for $\alpha=1.41$ and $\mu=3\;fm^{-1}$ has a lower temperature
parameter than that of $\alpha=0.47$ and $\mu=1\;fm^{-1}$. This is because more
work is been done when there are more relatively large angle scatterings and we
get more cooling. Scheme (b) and (c) can not get frame independent results
either.

\begin{figure}[h]
\vspace{1.0cm}
\hspace{0.01cm}
\psfig{figure=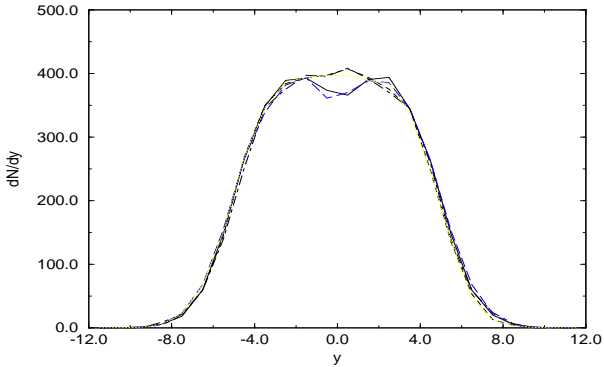,height=2.1in,width=1.9in,angle=-90} 
\vspace{-0.5cm}
\caption{
20 event averaged $\f{dN}{dy}$ distribution for scheme (a). 
The dot-dashed curve is for the initial distribution; 
the thick solid is for $\alpha=0.47$ and $\mu=1\;fm^{-1}$ collider frame
ordering;
the thick dashed curve is for $\alpha=0.47$ and $\mu=1\;fm^{-1}$ target frame
ordering;
the long dashed curve is for $\alpha=1.41$ and $\mu=3\;fm^{-1}$ collider frame
ordering;
the dotted curve is for $\alpha=1.41$ and $\mu=3\;fm^{-1}$ target frame
ordering. 
The dot-dashed, thick dashed and dotted curves overlap; the thick solid and
long dashed curves overlap.
}
\end{figure}

\begin{figure}[h]
\vspace{1.5cm}
\hspace{0.01cm}
\psfig{figure=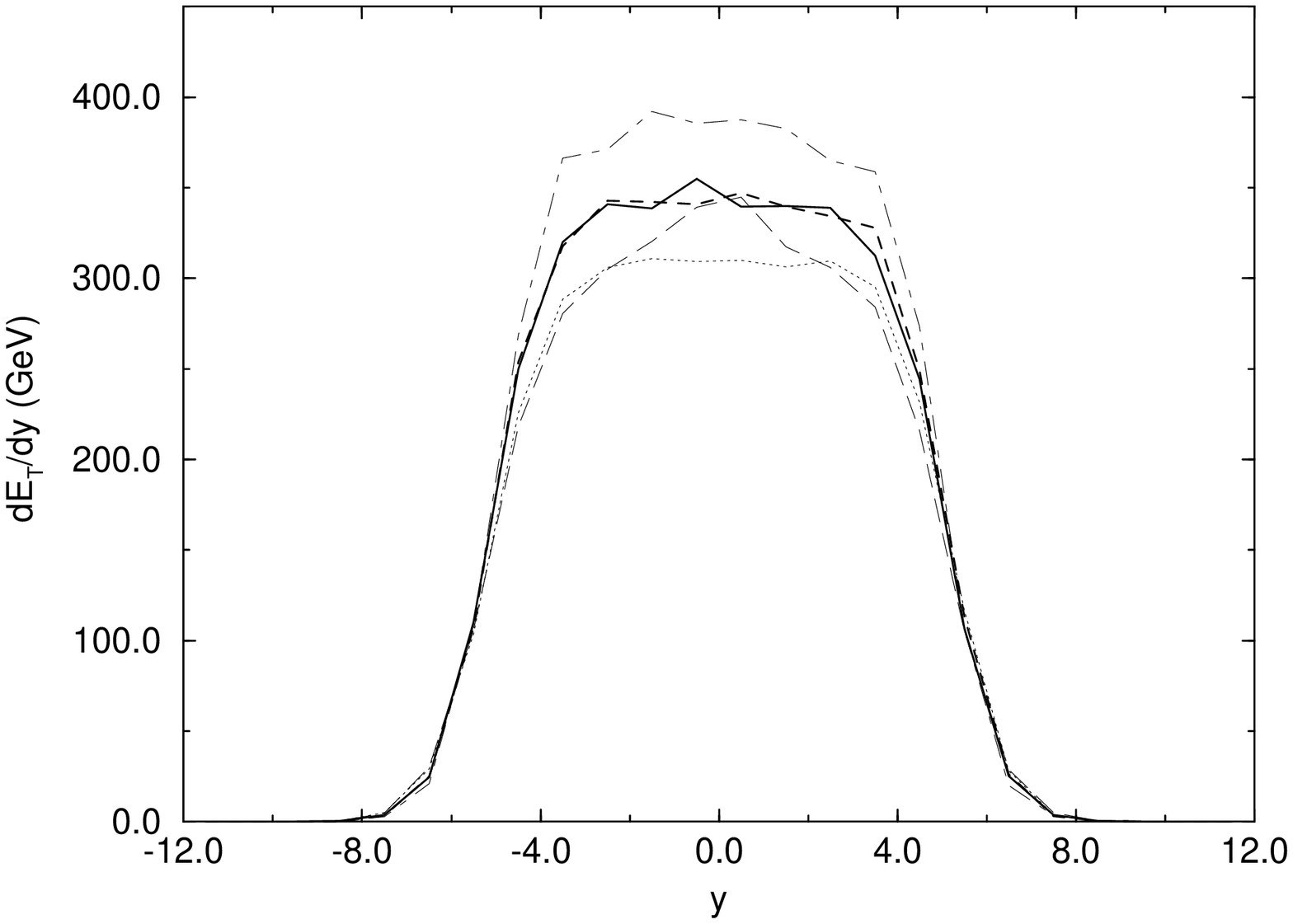,height=2.1in,width=1.9in,angle=-90} 
\vspace{-0.5cm}
\caption{
20 event averaged $\f{dE_T}{dy}$ distribution for scheme (a).
The dot-dashed curve is for the initial distribution; 
the thick solid is for $\alpha=0.47$ and $\mu=1\;fm^{-1}$ collider frame
ordering;
the thick dashed curve is for $\alpha=0.47$ and $\mu=1\;fm^{-1}$ target frame
ordering;
the long dashed curve is for $\alpha=1.41$ and $\mu=3\;fm^{-1}$ collider frame
ordering;
the dotted curve is for $\alpha=1.41$ and $\mu=3\;fm^{-1}$ target frame
ordering. 
}
\end{figure}

\begin{figure}[h]
\vspace{1.5cm}
\hspace{0.01cm}
\psfig{figure=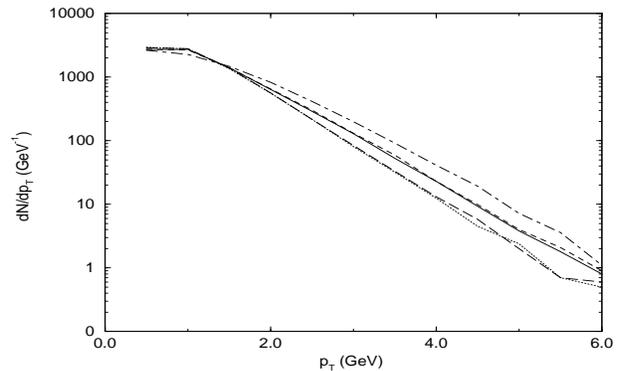,height=2.1in,width=1.9in,angle=-90} 
\vspace{-0.5cm}
\caption{
20 event averaged $\f{dN}{dp_T}$ distribution for scheme (a).
The dot-dashed curve is for the initial distribution; 
the thick solid is for $\alpha=0.47$ and $\mu=1\;fm^{-1}$ collider frame
ordering;
the thick dashed curve is for $\alpha=0.47$ and $\mu=1\;fm^{-1}$ target frame
ordering;
the long dashed curve is for $\alpha=1.41$ and $\mu=3\;fm^{-1}$ collider frame
ordering;
the dotted curve is for $\alpha=1.41$ and $\mu=3\;fm^{-1}$ target frame
ordering. 
}
\end{figure}

To study the effect of different differential cross sections, we compare the
results for $\alpha=0.47$ and $\mu=3\;fm^{-1}$ and $\alpha=1.41$ and
$\mu=9\;fm^{-1}$. Shown in Fig.~10 and Fig.~11 are $dE_T/dy$ and $dN/dp_T$
distributions. $dN/dy$ distribution is the same for these two cases and they
are the same as the
initial. From Fig.~10, a clear drop in the height of central rapidity plateau
can be seen with more large angle scatterings. This shows that more work has
been done and is also the reason that we get more cooling in $dN/dp_T$
distribution. We get $7,000$ collisions in the collider frame and $6,700$ in
the target frame. Scheme (b) gives the similar results and they are all frame
independent.

\begin{figure}[h]
\vspace{1.0cm}
\hspace{0.01cm}
\psfig{figure=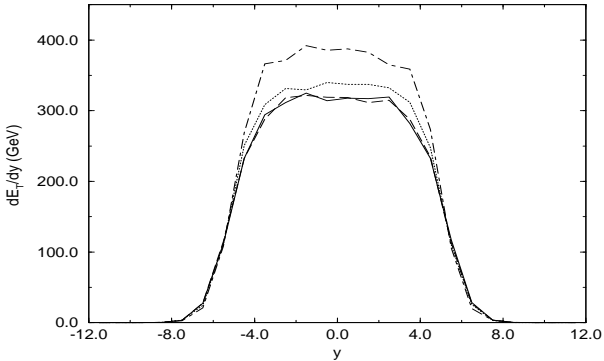,height=2.1in,width=1.9in,angle=-90} 
\vspace{-0.5cm}
\caption{
20 event averaged $\f{dE_T}{dy}$ distribution for scheme (a).
The dot-dashed curve is for the initial distribution; 
the dotted curve is for $\alpha=0.47$ and $\mu=3\;fm^{-1}$
the thick solid is for $\alpha=1.41$ and $\mu=9\;fm^{-1}$ collider frame
ordering;
the thick dashed curve is for $\alpha=1.41$ and $\mu=9\;fm^{-1}$ target frame
ordering.
}
\end{figure}

\begin{figure}[h]
\vspace{1.0cm}
\hspace{0.01cm}
\psfig{figure=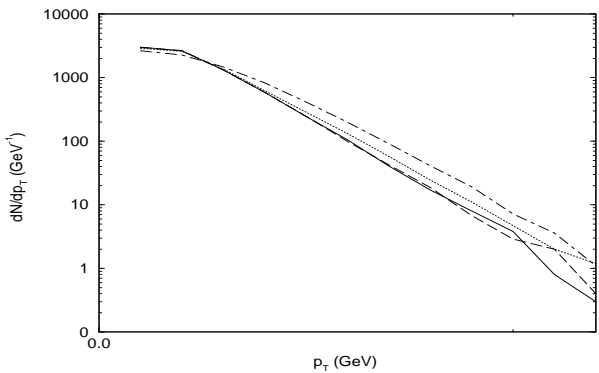,height=2.1in,width=1.9in,angle=-90} 
\vspace{-0.5cm}
\caption{
20 event averaged $\f{dN}{dp_T}$ distribution for scheme (a).
The dot-dashed curve is for the initial distribution; 
the dotted curve is for $\alpha=0.47$ and $\mu=3\;fm^{-1}$
the thick solid is for $\alpha=1.41$ and $\mu=9\;fm^{-1}$ collider frame
ordering;
the thick dashed curve is for $\alpha=1.41$ and $\mu=9\;fm^{-1}$ target frame
ordering.
}
\end{figure}

Another interesting thing to look at is the number of non-causal collisions
vs. the number of total collisions. A non-causal collision here is defined to
be a collision that the global ordering is not consistent with the ordering for
a particular particle participating in the collision. This is better seen in
scheme (b). If
particle a's next collision according to global ordering is with i, but the
collision time for particle a with i (i.e., the time a is going to change its
momentum) is later than the collision time for a with j (even though according
to global ordering, a j collision happens later), then we call this collision a
non-causal collision. For scheme (b) in the collider lab frame, with
$\alpha=0.47$ and $\mu=3\;fm^{-1}$, 
we get a total of $6,800$ collisions per event with $900$ of them
non-causal. With a larger cross section ($\alpha=0.47$ and $\mu=1\;fm^{-1}$),
as we expect, the non-causal to total ratio goes up. There are total $21,000$
collisions with $10,400$ of them non-causal. Boost to the target frame, the
total number of collisions for $\alpha=0.47$ and $\mu=3\;fm^{-1}$ is $6,500$
while the number of non-causal is $1,000$. The effects of putting in a mass
for the particles are also studied. There are no significant changes in
$\f{dE_T}{dy}$, $\f{dN}{dy}$ and $\f{dN}{dp_T}$ and the number of total and
non-causal collisions for massless, $m=0.01\;GeV$ and $m=0.1\;GeV$. Fig.~12
shows the collisions rates. We see that the results are not sensitive to small
masses of the particles. 

\begin{figure}[h]
\vspace{1.5cm}
\hspace{0.01cm}
\psfig{figure=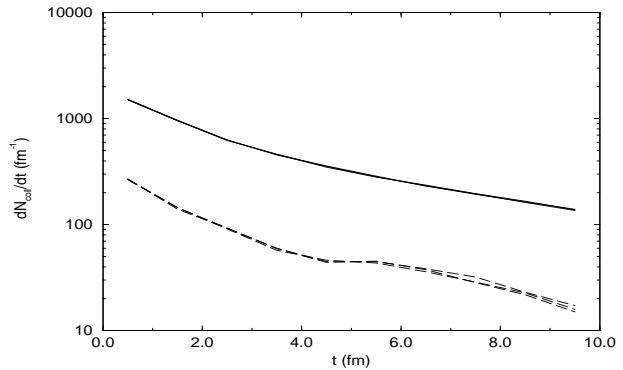,height=2.1in,width=1.9in,angle=-90} 
\vspace{-0.5cm}
\caption{
20 event averaged collision rate as a function of time for scheme (b). The
upper 3 solid curves are for massless, $m=0.01GeV$ and $m=0.1GeV$ total
collision rate and the lower 3 dashed curves are for non-causal collision rate
of these 3 masses.
}
\end{figure}

What we describe here is the parallel ensemble method in which average is taken
over $n$ independent events. An alternative, the full ensemble method
\cite{welke1}, is to
increase the particle number 
by $k$ (by dividing the particles into small pieces) and at the same time
decrease the cross section by a factor of $k$. 
In the full ensemble method, the
mean free path is the same as for the parallel ensemble method, but the cross
section is smaller and hence isolated 2 body scattering condition can be
satisfied. This makes it possible to eliminate the causality violation in the
limit of large $k$. 
But the full ensemble method will
increase the computation time significantly due to the increasing number of
checks required for the higher density system. 
We see from the above study that for reactions with strong inside outside
correlations, the full ensemble method may not have to be used in practice to
get reasonable results as long as a sensible scattering prescription is used.

\section{CONCLUSIONS}
After studying the frame dependence of parton cascade results, we see that
different schemes may give quite different results. Especially, the global
frame collision and ordering scheme results have a strong frame dependence. 
Other 2 particle center of mass frame collision and global frame ordering
schemes can give almost frame independent results even when the interaction
range is on the order of the mean free path. As we expect, they fail when the
interaction is much large than the mean free path. Different differential cross
sections lead to different longitudinal work and give different $dE_T/dy$ and
$dN/dp_T$ results. Since radiation will increase the cross section and parton
density, the quantum statistics will also change effective cross sections,
further study is necessary when we apply this to more realistic
situations.

\acknowledgments

We thank M. Asakawa, S. Gavin, K. Geiger, E. Gonzalez-Ferreiro, D. Kahana,
S. Kahana, Z. Lin,
R. Mattiello, C. Noack, D. Rischke, J. Randrup, R. Vogt, K. Werner for 
useful discussions, and M. Gyulassy for continuing encouragement and
discussions and a critical reading of the manuscript. 
B. Z. gratefully acknowledges partial support from the [Department of Energy]
Institute for Nuclear Theory at the University of Washington program INT-96-3
during the completion of this work. This work
was supported by the U.S. Department of Energy under Contract
No. DE-FG02-93ER40764 and DE-FG-02-92ER40699. 

%----------------------------References---------------------------------
{}

\begin{thebibliography}{20}

\bibitem{ypang1}
Y. Pang, D. E. Kahana, S. H. Kahana and T. J. Schlagel, {\it Nucl. Phys. A}
{\bf 590} (1995) 565c; 

H. Sorge, H. St\"oker and W. Greiner, {\it
Nucl. Phys. A} {\bf 498} (1989) 567c; 

K. Geiger, {\it Phys. Rev. D} {\bf 46} (1992) 4965, 4986.

\bibitem{peter1}
e.g.,
T. Kodama, S. B. Duarte, K. C. Chung, R. Donangelo and R. A. M. S. Nazareth,
{\it Phys. Rev. C} {\bf 29} (1984) 2146;
G. Peter, D. Behrens and C. C. Noack, {\it Phys. Rev. C} {\bf 49} (1994) 3253.

\bibitem{kortmeyer1}
e.g. see G. Kortmeyer, W. Bauer, K. Haglin, J. Murray and S. Pratt, {it
Phys. Rev. C} {\bf 52} 2714.

\bibitem{zhang1}
Further details will be published later.

\bibitem{ruus1}
P. V. Ruuskannen, {\it Phys. Lett.} {\bf 147 B} (1984) 465; 

K. J. Eskola and M. Gyulassy, {\it Phys. Rev. C} {\bf 47} (1993) 2329;

B. Zhang, M. Gyulassy and Y. Pang, CU-TP-795.

\bibitem{welke1}
see e.g., G. Welke, R. Malflied, C. Gregoire, M. Prakash and E. Suraud, {\it
Phys. rev. C} {\bf 40} (1989) 2611; Kortemeyer et.al. cited above.

\end{thebibliography}
\end{document}